\documentclass[mathleft
]{an}
\usepackage{graphicx}
\usepackage{times}
\begin{document}

% The following seven commands are intended for editorial usage and should be 
% the author(s).
\Pagespan{789}{}% Document's page range.
% If second parameter is left empty, the last page is computed automatically.
\Yearpublication{2006}%
\Yearsubmission{2005}%
\Month{11}%
\Volume{999}%
\Issue{88}%
% \DOI{This.is/not.aDOI}%

%\lhead[\thepage]{L. \v C. Popovi\'c et al.:
% Amplification and variability of the AGN X-ray emission due to microlensing}
% \rhead[Astron. Nachr./AN~{\bf XXX} (2004) X]{\thepage}
%\headnote{Astron. Nachr./AN {\bf 32X} (2004) X, XXX--XXX}

\title{Amplification and variability of the AGN X-ray emission due to 
microlensing}

\author{L.\v C. Popovi\'c\inst{1}\fnmsep\thanks{Corresponding author:
  \email{lpopovic@aob.bg.ac.yu}\newline}, P. 
Jovanovi\'c\inst{1}, T. 
Petrovi\'c\inst{1,2}, V. N. Shalyapin\inst{3}}

\titlerunning{Amplification and variability of the AGN X-ray emission due to 
microlensing}
\authorrunning{L. \v C. Popovi\'c et al.}

\institute{Astronomical Observatory, Volgina 7, 11160 Belgrade
74, Serbia
\and
Department of Astronomy, Faculty of Mathematics, University of
Belgrade,  Studentski trg 16, 11000  Belgrade, Serbia
\and 
 Usikov Institute of Radiophysics and  Electronics, 12 Proskura st., 
Kharkov, 61085 
Ukraine}

\received{20 Avg 2006}
\accepted{11 Oct 2006}
\publonline{later}


\abstract{We consider the contribution of  microlensing to the AGN Fe 
K$\alpha$ line and 
X-ray continuum amplification and variation. To
investigate the variability of the line and X-ray continuum, we  studied
the effects of microlensing on quasar X-ray spectra produced by  crossing
of a microlensing pattern across a  standard relativistic accretion disk.
To describe the disk emission we used a ray tracing method
considering both metrics, Schwarzschild and Kerr.
We found that the Fe K$\alpha$ and  continuum may experience
  significant   amplification by a microlensing event (even for 
microlenses of very 
small mass). 
  Also, we investigate a contribution of microlensing to the X-ray 
variability of 
 high-redshifted  QSOs, finding  that cosmologically distributed 
 deflector  may contribute significantly to the X-ray variability of 
high-redshifted 
QSOs (z$>$2).
    }

%\correspondence{lpopovic@aob.bg.ac.yu}

\maketitle

\section{Introduction}

 The X-ray emission in Active Galactic Nuclei -- AGN  (the continuum as well as in 
 the Fe K$\alpha$ spectral line) has rapid and irregular variability (see e.g. 
 Manners et al. 2002). 
 X-ray flux variability has long been known to be a common property of AGN, i.e. 
 X-ray flux variations are observed on timescales from $\sim$1000 s to years, and 
 amplitude variations of up to an order of magnitude are observed in the 
 $\sim$0.1-10 keV spectral band (see, for example
 reviews by Mushotzky et al. 1993; Ulrich et al. 1997, and references therein).

Recent observational  studies suggest that gravitational
microlensing can induce variability in the X-ray emission of
lensed QSOs. Microlensing of the Fe K$\alpha$ line
has been reported at least in three macrolensed QSOs:  MG
J0414+0534 (Chartas et al. 2002),  QSO 2237+0305
(Dai et al. 2003), and H1413+117
(Popovi\'c et al. 2003a, Chartas et al. 2004).

The influence of microlensing in the X-ray emission has been also
theoretically investigated. Mineshiga et al. (2001) simulated the 
variation of the
X-ray continuum due to microlensing showing that the flux magnifications
for the X-ray and optical continuum emission regions are not
significantly different during the microlensing event, while
Yonehara et al. (1998,1999), Takahashi et al. (2001) found that simulated 
spectral variations caused
by
microlensing show different behavior, depending on photon energy. Also,
microlensed light curves for thin accretion disks around Sch\-war\-zschild and
Kerr black holes were considered in Jar\-os\-zy\'ns\-ki (2002) and microlensing 
light
curves for the Fe K$\alpha$ were simulated by Jar\-os\-zy\'n\-ski (1992). 
Moreover, the 
influence of microlensing in the Fe K$\alpha$ spectral line
shape was discussed in Popovi\'c (2001a), Chartas (2002),
Popovi\'c et al. (2003ab, 2006) and Jovanovi\'c (2005,2006).

All of these investigations showed that
 monitoring of
gravitational lenses may help
us to understand the physics  of the innermost part of AGNs, i.e. the physics of
 relativistic accretion  disks. The aims of this work are to discuss of the X-ray 
 variation due to microlensing (in the X-ray continuum and  Fe K$\alpha$ 
  line) and consider the probability that the QSO X-ray emission  
  may be affected by microlensing of cosmologically distributed objects. 

\section{Theoretical model}

%\subsection{X-ray emitting region geometry}
\noindent{\bf The X-ray emitting region geometry.}

According to the standard model of AGNs, a QSO consists of a black
hole (BH)
surrounded by a (X-ray and optical) continuum emitting region probably
with an accretion disk geometry,
a broad line region and a larger region that can be resolved in several
nearby
AGN that  is the so called  narrow line region
(e.g., Krolik 1999).  The X-ray emitting region is supposed to be the most compact 
and the closest to the  massive black hole. Consequently,
an initial assumption  that we adopted was the existence of a super-massive BH  
(10$^7$-10$^9$ M$\odot$) surrounded by an accretion disk that radiates X-rays in
the center of all types of  AGNs. Accretion disks could have different forms, 
dimensions, and emission,  depending on the type of central BH, whether it can be 
rotating (Kerr metric) or non-rotating (Schwarzschild metric) BH. Except for 
effects due to disk instability,  its emission could also be affected by 
gravitational microlensing,  especially in the case of gravitationally lensed 
QSOs (Chartas et al. 2002,2004; Dai et al. 2003).

     The disk emission was analyzed using numerical simulations based on a
ray-tracing method in a Kerr metric, taking into account only photon
trajectories reaching the observer's sky plane (see Popovi\'c et al. 2003ab and references therein).
The assumption of a disk geometry for the distribution of  the Fe
K$\alpha$ emitters is supported by the spectral shape of this line in
AGN (e.g. Nandra 1997). Regarding the
X-ray continuum emission, it seems
that it mainly arises from an accretion disk. For instance,
Fabian et al. (2003) have shown that the X-ray spectral variability of
MCG-6-30-15 can be modeled  by a two-component model where the one varying
 component is a power-law and the other constant component is produced
by very strong reflection from a relativistic disk.
Consequently, to study the effects of microlensing  on a compact accretion disk we 
 used the ray tracing method considering only those photon
trajectories that reach the sky plane at a given observer's angle
$\theta_{\rm  obs}$ (see e.g. Popovi\'c et al. 2003ab and references therein). The
amplified  brightness with amplification $A(X,Y)$ for the
 continuum and the line is given in Popovi\'c et al. (2006) as well as 
discussion about emissivity of the disk, and here will not be repeated.

\vspace{.1cm}
%\subsection{Microlens model}
\noindent{\bf Microlens model.}

 The influence of microlensing  on a standard accretion disk was studied 
using three 
 types of a microlensing model:  po\-int-li\-ke microlens, st\-rai\-ght-fold 
ca\-us\-tic, and 
quadrupole microlens.  Illustration of the 
point like 
microlens and 
 st\-rai\-ght-fold  caustic crossing over an accretion disk in the 
Kerr/Sch. 
 metric and the  corresponding effects on the  the X-ray 
continuum shapes and the 
 Fe K$\alpha$ line can be found in Popovi\'c et al. (2003ab, 2006) and 
Jovanovi\'c 
(2005,2006).

For the each lens system we are able to model the
amplification maps  (see e.g. Abajas et al.
2005, Popovi\'c et  al. 2006 and Figs. 1 and 2). In order to apply an appropriate 
microlens model,    we  considered a standard microlensing
magnification pattern (Figs. 1 and 2, upper)
for different objects. 
The simulation was made employing ray-shooting techniques that send rays
 from the observer through the lens to the source plane (Kayser et al.
 1986; Schneider \& Weiss 1987; Wambsganss et al 1990a,b). We assume a
flat cosmological model with $\Omega
=0.3$  and $H_{o}= 75\ \rm km\ s^{-1} Mpc^{-1}$.

As an example, the  illustration of the amplification patterns for PG 
1115+080A1,A2 images  are given in Figs. 1 and 2 (upper panel). The 
dimensions of the patterns are
 16$\times$ Einstein  radii (ERR) on a side. $\kappa$=0.56,
$\lambda$=0.11 and $\kappa$=0.63, $\lambda$=0.11 are taken for A1
and A2, respectively ($\kappa_c$=0). The  mass of microlens is
taken to be 1$M_\odot$.

\begin{figure}
\includegraphics[width=7.4cm]{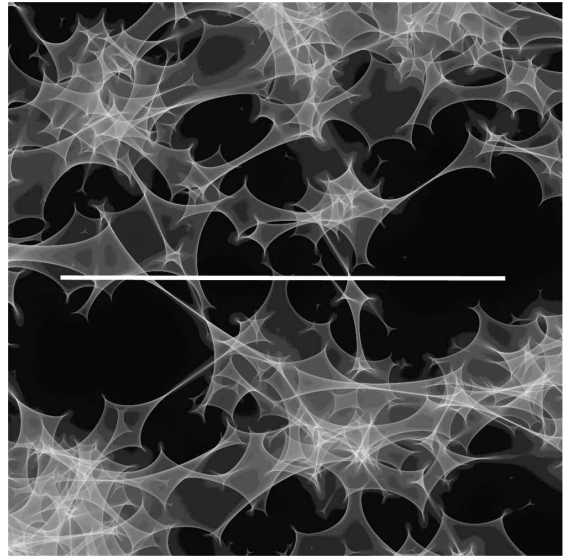}
\includegraphics[width=7.4cm]{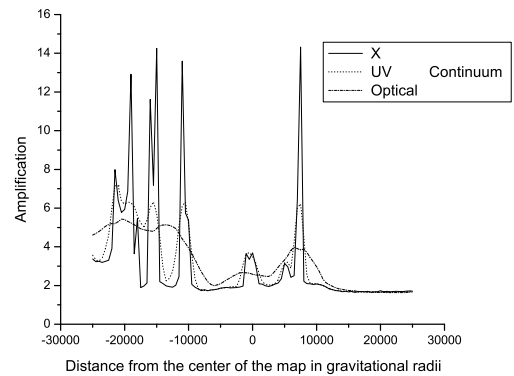}
\caption{{\it Top}: The magnification map of the PG1115+080A1
image. The white solid line represents the center position of an
accretion disk. {\it Down}: Corresponding variation in the X-ray
(solid), UV (dotted) and optical (dash-dotted) spectral band.}
\end{figure}

\begin{figure}
\includegraphics[width=7.4cm]{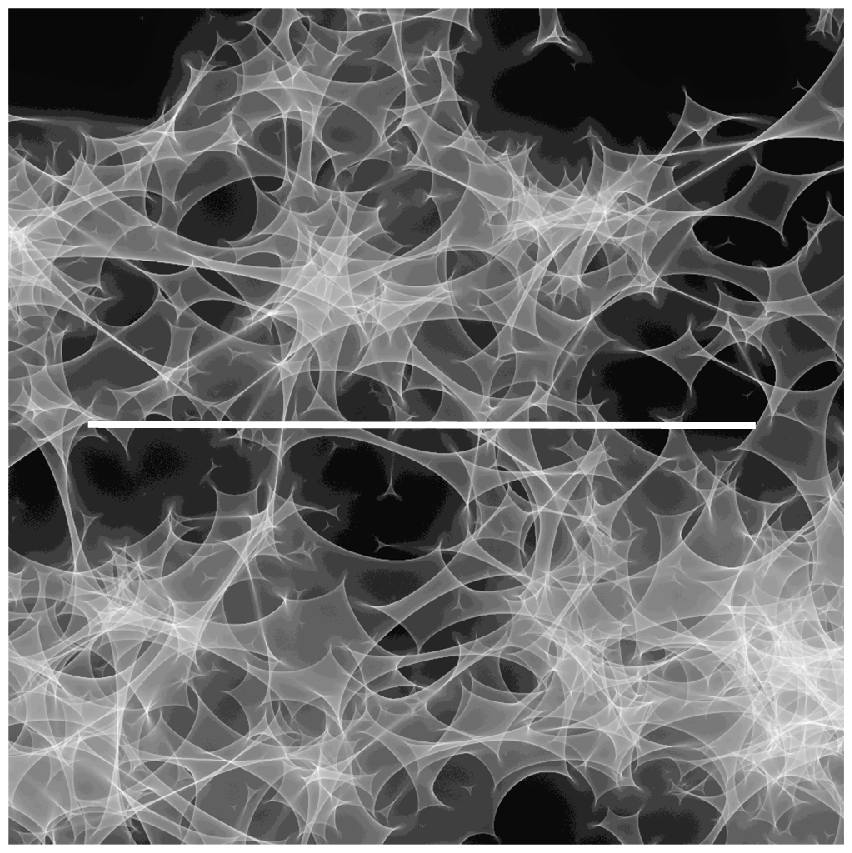}
\includegraphics[width=7.4cm]{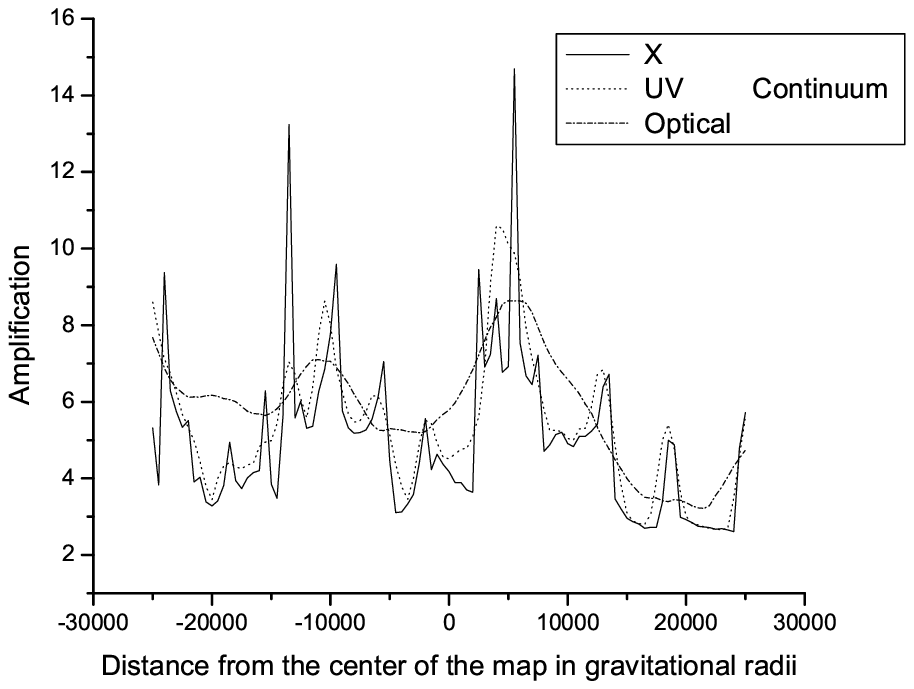}
\caption{The same as in Fig. 1, but for the PG1115+080A2 image.}
\end{figure}

\section{Probability of the X-ray emission microlensing}

The optical depth  $\tau$ is the chance of seeing a microlens, i.e. the 
probability 
that at any instant  of time a source is covered by the Einstein ring of a 
deflector. It was  shown that deflectors from the host bulge  and halo 
have a small 
optical depth   $\tau\sim 10^{-4}-10^{-3}$ (Popovi\'c et al. 2003a, 
Zakharov et al. 
2004,2005).  On the other hand, Zakharov et al. (2004,2005) showed that 
cosmologically   distributed objects can have high optical depth (for 
the high 
red-shifted QSOs it may be $\tau\sim 0.1$)

Moreover, one  should take into account that, as it was mentioned above, 
the X-ray 
emission 
comes from  the most compact part of the accretion disk. To demonstrate the 
differences  in light curve (or amplification) of different spectral bands we 
investigate   microlensing of the X-ray, UV and optical emission assuming 
that the
source mass is $10^8\ M_\odot$. For modeling of emitting regions  an 
accretion disk in 
Sch\-wa\-rzs\-child metric is used. The radii 
of emitting
regions are taken as: $R_{in}= 500\ R_g$, $R_{out}=5000\ R_g$ -- for the 
optical
emitting region; $R_{in}= 100\ R_g$, $R_{out}=1000\ R_g$ -- for the UV
emitting region; and  $R_{in}= R_{ms}$\footnote{$R_{ms}$ is the radius of 
the 
marginal stability orbit}, $R_{out}=50\ R_g$ -- for the X-ray emitting 
region.  For amplification patterns 
we use ones  calculated for  PG 1115+080A1,A2 (Figs. 1 and 2, upper panel). The 
variation in  total flux for all three spectral bands is given in Figs. 1 and 2 
(lower panel). The total fluxes are integrated in next spectral bands:
optical, from  3500 to 7000 \AA\  (dashed line); UV, from 1000 to 3500 
\AA\ (dotted 
line); 
and X-ray, from 1.24 to 12.4 \AA \ (i.e. 1-10 KeV -- solid line).

As one can see from Figs. 1 and 2, the variation in the  X-ray is higher than in 
the UV/optical spectral band. Also, one can  see that during one microlensing event 
of 
the optical emission region, one can expect several microlensing events in the 
X-ray  spectral band. It is expected, because
the  X-ray emission regions are much smaller than UV/optical ones, and even small 
mass  deflectors from a foreground galaxy can produce significant magnification in 
the  X-ray radiation (Popovi\'c et al. 2003a), while it will not happen in the 
UV/optical  band. Consequently, this is an additional factor that should be taken 
into account  in the calculation of the X-ray emission optical depth.

 The probability that the shape of the Fe K$\alpha$ line is distorted (or 
amplified) is highest in gravitationally lensed systems.
 The probabilities that  lensed QSOs, where microlensing of the Fe K$\alpha$ were 
 observed (QSO H1413+117, QSO 2237+0305  and J0414+0534), are gravitationally 
 mi\-cro\-len\-sed by objects in a foreground galaxy  and by 
 cosmologically distributed objects  are calculated by Zakharov 
  et al. (2004) (see their Table 2). It is interesting that the optical 
depth for 
  microlensing by cosmologically distributed microlenses can be one order 
higher than 
for microlensing by objects in a foreground galaxy. So the observed microlensing in  
the X-ray, e.g. Fe K$\alpha$ line, from these objects might be caused by 
cosmologically  
distributed objects rather than by the objects from a lens galaxy.

 Taking into account this result and results from our simulations (Figs. 1 and 2) it 
seems that there is a big chance to  find X-ray microlensing for  
gravitationally lensed systems that have signatures of microlensing  in the 
UV/optical band. Moreover, considering the sizes of the sources of X-ray  
radiation, the 
variability in the X-ray range during microlensing event is more  prominent than in 
the optical and UV band (see Figs. 1 and 2). Consequently,  gravitational 
microlensing in the X-ray band is a powerful tool for dark  matter investigations, 
as the upper limit of optical depth ( $\tau\sim 0.1$)  calculated by Zakharov et 
al. (2004,2005) corresponds to the case where dark matter  forms cosmologically 
distributed deflectors.

From discussion above and our simulations,  it seems that contribution of the X-ray 
radiation microlensing can have a significant part in the X-ray variations of 
lensed 
and un-lensed high red-shifted QSOs.

\section{Discussion}

Recent observations of three lensed QSOs seem to support  idea about the X-ray  
microlensing (Chartas et al. 2002;  Dai et al. 2003).
Also, it seems that the Fe K$\alpha$ line has been more affected by microlensing  
than the X-ray continuum. Popovi\'c et al. (2003a,2006) showed that objects in a 
foreground galaxy with very small  masses can cause strong cha\-n\-ges in the 
X-ray 
line profile. Our investigations  indicate that the observational probability of 
X-ray variation due to microlensing  events is higher than in the UV and optical 
radiation of QSOs. This is connected  with the fact that typical sizes of X-ray 
emission regions are much smaller than  typical sizes of those producing optical 
and UV bands. Typical optical and UV  emission region sizes could be comparable or 
even larger than Einstein radii of  microlenses and therefore microlenses magnify 
only a small part of the region  emitting in the optical or UV band (see e.g. 
 Abajas et al. 2002,2005; Popovi\'c et al.  2001b, for UV and optical 
spectral line 
region). This is reason that it could  be a very tiny effect in UV/optical spectral 
band from an observer point of view.  On the other hand, the variability and 
amplification of the X-ray spectra seems  to be more prominent and it may be 
detectable. Note here, that expected microlensing time scales for X-ray microlensing  are 
from
several days to several months  (e.g. from our X-ray microlensing simulations for
several
microlensed  QSOs, we obtained the  shortest time scale for Q2237+0305 of
$\approx$ 14 days,  while for LBQS 1009-0252 we obtained  the longest one
$\approx$ 320 days). Heaving in mind that the variation of optical/UV continua, which are 
weaker and
much slower (the order of several years, see Fig. 2 in Jovanovi\'c 2006), one
can expect that in a period of several years (during microlensing of the optical
emission) a number of microlensing of the X-ray emission will occur. But, after a
complete microlensing of the optical emission region, there should be a
correlation between X-ray and optical emission amplification (see Figs. 1 and 2).

From theoretical  investigation  of the Fe  K$\alpha$ line microlensing  
there are 
several interesting results (see Popovi\'c et al. 2003ab):

i) Microlenses of very small projected  Einstein radii ($\sim $ 10 $R_{\rm g}$) 
can 
give rise to significant changes in the  iron line profiles. The effects may be an 
order of magnitude greater than the ones  inferred for the UV and optical lines. 
Off-centered microlenses would induce  strong asymmetries in the observed 
Fe K$\alpha$ line 
profiles. In the case that a part of the Fe K$\alpha$ line  originates from the
reflection on a distant matter, the amplification will be present only in the
broad component from the disk, and the effects of microlensing  will  be
similar as in the case of a pure disk emitted line.

ii) The effects of microlensing show  differences in the Kerr and Schwarzschild 
metrics, the amplitude of the magnification  being greater in the Kerr metric. The 
transit of a microlens along the rotation  axis of the accretion disc would induce 
a strong amplification of the blue peak in  the Schwarzschild metric when the 
microlens was centered in the approaching part.  In the Kerr metric the 
amplification will be greater but will not affect  so preferentially the blue part 
of the line. This difference could be interesting  to probe the rotation of an 
accretion disc.

\section{Conclusion}

 From recent investigation of the X-ray emitting region microlensing 
 (Popovi\'c et al. 2003ab, 2006, Zakharov et al. 2004,2005, Jovanovi\'c 2005,2006) 
and simulations per\-for\-med in this paper 
we give following 
conclusions:

   1. Gravitational microlensing can produce significant \break var\-iat\-ions 
and
amplifications of the line and continuum flu\-xes.  These deformations of the 
X-ray 
radiation depend on both, the disk and microlens parameters.

   2. Microlensing can satisfactorily explain the excess in the Fe K$\alpha$ line 
observed in three gravitational lens systems:  MG J0414+0534 (Chartas et al. 2002), 
QSO 2237+0305 (Dai et al. 2003), and H1413+117 (Chartas et al. 2004).

   3.  On the basis of these investigations, one can expect that the Fe K$\alpha$
line
and X-ray continuum  amplification due to microlensing can be significantly larger 
than the corresponding  effects on optical/UV emission lines and continua.

4. The optical depth  for gravitational microlensing by cosmologically distributed 
deflectors could be  significant. The 
maximum optical depth  could  be expected if dark matter forms cosmologically 
distributed compact objects.

Results mentioned above and recent   investigation of the X-ray microlensing 
show that monitoring of lensed and un-lensed  high-redshift QSOs in X-ray band 
can be very useful 
not only for investigation of the innermost structure of QSOs, but also 
 of the cosmological parameters.

\begin{acknowledgements}
This work is a part of the project P146002  ``Astrophysical Spectroscopy of
Extragalactic Objects'' supported by the
Ministry of Science
of Serbia.
\end{acknowledgements}

{}


\begin{thebibliography}{}

\bibitem[]{}
Abajas, C., Mediavilla, E.G., Mu\~noz, J.A., Popovi\'c, L. \v C.,
 Oscoz A. 2002,  ApJ 576, 640.

\bibitem[]{}
Abajas, C., Mediavilla, E.G.,  Mu\~noz, J.A., Popovi\'c,
L. \v C. 2005,  MemSAIt Suppl. 7, 48



\bibitem[]{}
Chartas, G., Agol, E., Eracleous, M., Garmire, G., Bautz, M. W., Morgan,
N. D. 2002,  ApJ,  568, 509.

\bibitem[]{}
Chartas, G., Eracleous, M., Agol, E., Gallagher, S. C. 2004, ApJ,
606, 78.

\bibitem[]{}
Dai, X., Chartas, G., Agol, E., Bautz, M. W., 
  Garmire, G.P. 2003, ApJ, 589, 100.

%\bibitem[]{}
%Dai, X., Chartas, G., Eracleous, M. 
%  Garmire, G.P. 2004, ApJ, 605, 45.




\bibitem[]{}
Fabian, A.C., Vaughan S.: 2003,  MNRAS, 340, L28.

\bibitem[]{}
Jaroszy\'nski, M.: 2002, Acta Astronomica, 52, 203.

\bibitem[]{}
Jaroszy\'nski, M., Wambsganss, J.W., Paczy\'nski, B.: 1992, ApJ, 396, L65.

\bibitem[]{}
Jovanovi\'c, P.: 2006 MemSAIt Suppl., 7, 56

\bibitem[]{}
Jovanovi\'c, P.: 2006 PASP, 118, 656

\bibitem[]{}
Kayser, R., Refsdal, S.,  Stabell, R.: 1986, A{\&}A, 166, 36.

\bibitem[]{}
Krolik, J. H.: 1999, Active Galactic Nuclei, Princeton University Press, New Jersey.

\bibitem[]{}
Lawrence, A.,  Papadakis, I.:  1993,  ApJ,  414, 85

\bibitem[]{}
Mineshige, S., Yonehara, A.,  Takahashi, R.: 2001,
PASJ, 18, 186.

\bibitem[]{}
Manners, J., Almaini, O.,  Lawrence, A.: 2002,  MNRAS,  330, 390

\bibitem[]{}
Mushotzky, R. F., Done, C., Pounds, K. A.: 1993, 
 ARA\&A, 31, 717

\bibitem[]{}
 Nandra K., George I.M., Mushotzky R.F., Turner T.J. 
Yaqoob T.: 1997, ApJ. 477, 602.



\bibitem[]{}
Popovi{\'c}, L., \v C.,  Mediavilla, E.G., Mu\~noz, J.,
Dimitrijevi\'c, M.S.,  Jovanovi\'c, P.: 2001a,  SerAJ,
164, 73.

\bibitem[]{}
Popovi{\'c}, L.\v C.,  Mediavilla, E.G.,  Mu\~noz, J.: 2001b,
A{\&}A 378, 295.

\bibitem[]{}
 Popovi\'c, L.\v C., Mediavilla, E.G., Jovanovi\'c, P.,  Mu\~noz, J.A.:
 2003a,  A{\&}A, 398, 975.

\bibitem[]{}
Popovi\'c, L.\v C., Jovanovi\'c, P., Mediavilla, E.G.,  Mu\~noz,
J.A.: 2003b, Astron.  Astrophys. Transactions, 22, 719.

\bibitem[]{}
Popovi\'c, L. \v C., Jovanovi\'c, P.,  Mediavilla, E., Zakharov, A. F., Abajas, 
C., Muoz, J. A., Chartas, G.: 2006, ApJ, 637, 620

\bibitem[]{}
Schneider, P.,  Weiss, A.: 1987, A{\&}A, 171, 49.


\bibitem[]{}
Takahashi, R., Yonehara, A., Mineshige, S.: 2001, PASJ
53, 387.



\bibitem[]{}
Ulrich, M.-H., Maraschi, L.,  Megan, C.M.: 1997, ARA\&A, 35, 445


\bibitem[]{}
Wambsganss, J., Paczynski, B.,  Katz, N.: 1990a, ApJ, 352, 407.

\bibitem[]{}
Wambsganss, J., Schneider, P.,  Paczynski, B.: 1990b, ApJ, 358, L33.


\bibitem[]{}
Yonehara, A., Mineshige, S.,  Fukue, J., Umemura, M., Turner,
E.L.: 1999,
A{\&}A, 343, 41.

\bibitem[]{}
Yonehara, A., Mineshige, S., Manmoto, T., Fukue, J., Umemura, M., Turner,
E.L.: 1998,
ApJ, 501, L41; ApJ, 511, L65.

\bibitem[]{}
Zakharov, A.F., Popovi\'c, L. \v C., Jovanovi\'c, P.:  2004,
 A{\&}A, 420, 881.

\bibitem[]{}
Zakharov, A.F., Popovi\'c, L. \v C., Jovanovi\'c, P.:
 2005, in Proc. of IAU Symp. 225,
%, Gravitational Lensing Impact on Cosmology, 
%ed. Y. Mellier \& G. Meylan (Cambridge: 
Cambridge Univ. Press, 363


\end{thebibliography}
\end{document}